\documentclass[11pt,a4paper]{article}
\pdfoutput=1
\usepackage{jcappub}

\usepackage{graphicx}
\usepackage{bm}
\usepackage{float}
\usepackage{subfigure}
\usepackage{amsmath}
\usepackage{amsfonts}
\usepackage{color}
\usepackage{slashed}
\usepackage{fancyhdr}

\newcommand{\GN}{G_{\rm N}}

\newcommand{\dd}{\text{d}}

\newcommand{\sg}{\sigma_g}

\begin{document}

\title{Ideal MHD(--Einstein) Solutions Obeying The Force-Free Condition}

\author{Yi-Zen Chu and Vitaly Vanchurin}
\affiliation{
Department of Physics and Astronomy, University of Minnesota, 1023 University Dr., Duluth, MN 55812, USA
}

\abstract{ We find two families of analytic solutions to the ideal magnetohydrodynamics (iMHD) equations, in a class of 4-dimensional (4D) curved spacetimes. The plasma current is null, and as a result, the stress-energy tensor of the plasma itself can be chosen to take a cosmological-constant-like form. Despite the presence of a plasma, the force-free condition -- where the electromagnetic current is orthogonal to the Maxwell tensor -- continues to be maintained. Moreover, a special case of one of these two families leads us to a fully self-consistent solution to the Einstein-iMHD equations: we obtain the Vaidya-(anti-)de Sitter metric sourced by the plasma and a null electromagnetic stress tensor. We also provide a {\sf Mathematica} code at \cite{MHDSystem} that researchers may use to readily verify analytic solutions to these iMHD equations in any curved 4D geometry.}

\maketitle

\section{Introduction and Setup}
\label{Section_Introduction}

Ideal magnetohydrodynamics (iMHD) is the electromagnetic system consisting of a perfect fluid plasma coupled to Maxwell's electromagnetic fields. Its ``force-free" limit, to be explained shortly, is believed to be relevant for describing the highly magnetized atmospheres of rapidly rotating neutron stars (aka pulsars) \cite{Pulsar,Goldreich:1969sb}. This ``Force-Free-Electrodynamics" (FFE) may also be applied to analyze the magnetized environment around supermassive spinning black holes \cite{Blandford:1977ds}, which through the Penrose process could explain astrophysical observations of energetic phenomenon such as quasars and highly relativistic jets from active galactic nuclei. 

In this paper, we embark on a program to seek iMHD solutions in curved geometries. As a starting strategy, we shall do so by extending the recently discovered FFE solutions in \cite{Brennan:2013kea,Gralla:2014yja} to the iMHD case. We also hope the code in \cite{MHDSystem}, which has aided us in obtaining the solutions reported here, will provide a useful tool for other researchers.

{\bf Setup} \qquad It is possible to encode the dynamics of iMHD in a 4D curved metric $g_{\mu\nu}$ through an action principle. The basic degrees of freedom are 3 scalar fields, which we will denote as
\begin{align}
\label{MHD_3Scalars}
\Phi^\text{I} = \Phi_\text{I} , \qquad\qquad \text{I} \in \left\{ 1,2,3 \right\} .
\end{align}
From the first two scalar fields we may build the Maxwell field strength
\begin{align}
F_{\mu\nu} \equiv \partial_{[\mu} \Phi^1 \partial_{\nu]} \Phi^2 ,
\end{align}
where the anti-symmetrization symbol is defined as $F_{[\mu\nu]} \equiv F_{\mu\nu} - F_{\nu\mu}$. From all three $\{ \Phi^\text{I} \}$ we may then form what we will call the plasma current\footnote{The $\widetilde{\epsilon}$ is the covariant Levi-Civita tensor; specifically, $\widetilde{\epsilon}_{\mu\nu\alpha\beta} = \sqrt{|g|} \epsilon_{\mu\nu\alpha\beta}$, where $\epsilon_{\mu\nu\alpha\beta}$ is the Levi-Civita symbol, which in turn returns the sign of the permutation that brings $\{ 0,1,2,3 \}$ to $\{ \mu, \nu, \alpha, \beta \}$ or is zero otherwise.}
\begin{align}
\label{PlasmaCurrent}
n^\mu 
&\equiv \widetilde{\epsilon}^{\mu \alpha\beta\gamma} \nabla_\alpha \Phi^1 \nabla_\beta \Phi^2 \nabla_\gamma \Phi^3  \\
\label{PlasmaCurrent_ContainsFmunu}
&= \frac{1}{2} \widetilde{\epsilon}^{\mu \alpha\beta\gamma} F_{\alpha\beta} \nabla_\gamma \Phi^3 ,
\end{align}
This plasma current (or the mass density current) $n^\mu$ describes the mass flow of both positively and negative charged particles. This is to be distinguished from the electromagnetic current that we will introduce below. The physical feature specific to magnetohydrodynamics is that the plasma current itself is orthogonal to the Maxwell tensor,
\begin{align}
n^\mu F_{\mu\nu} = 0 ;
\end{align}
this is an identity because $\nabla_\alpha \Phi^\text{I} n^\alpha = 0$ for all $\text{I} \in \{ 1,2,3 \}$. At this point, we have the necessary ingredients to write down the iMHD action principle. It reads
\begin{align}
\label{MHD_Lagrangian}
S_\text{iMHD} = S_\text{Plasma} + S_\text{Maxwell} ,
\end{align}
with
\begin{align}
\label{MHD_Lagrangian_Plasma}
S_\text{Plasma} 	&\equiv -\int \dd^4 x \sqrt{|g|} \rho_0\left[\sg n^2/2\right] ,  \qquad\qquad n^2 \equiv n_\mu n^\mu ,	\\
\label{MHD_Lagrangian_EM}
S_\text{Maxwell}	&\equiv - \frac{1}{2} \int \dd^4 x \sqrt{|g|} \varphi^2 ;
\end{align}
where
\begin{align}
\sg &= +1, \qquad\qquad \text{ if } \qquad\qquad \eta_{\mu\nu} = \text{diag}[+1,-1,-1,-1] \qquad \text{(``Mostly minus")} , \\
\sg &= -1, \qquad\qquad \text{ if } \qquad\qquad \eta_{\mu\nu} = \text{diag}[-1,+1,+1,+1] \qquad \text{(``Mostly plus")} .
\end{align}
and\footnote{The plasma action in eq. \eqref{MHD_Lagrangian_Plasma} does not take into account additional conserved charges that might be present; for a field theoretic discussion, see \cite{ParticleFluid}.}$^{,}$\footnote{Throughout this paper we will write our results in a metric sign convention independent manner, in order to make the results here and in the accompanying {\sf Mathematica} code accessible to readers from ``both sides of the aisle". For instance, $\sg n^2 > 0$ if $n^\mu$ is timelike in either sign convention.}
\begin{align}
\varphi^2 \equiv \frac{1}{2} F_{\mu\nu} F^{\mu\nu} .
\end{align}
We also remark that the action principle in eq. \eqref{MHD_Lagrangian} is in fact a special case of the Schubring-Vanchurin class of theories \cite{Schubring:2014ena},
\begin{align}
\label{SchubringVanchurin_Lagrangian}
S_{\text{Schubring-Vanchurin}} \equiv \int\dd^4 x \sqrt{|g|} \mathcal{L}\left[ \sg n^2/2, \varphi^2/2 \right]
\end{align}
describing the so-called {\it string fluids} which generalize both perfect particle fluids \cite{ParticleFluid} and pressure-less string fluids \cite{StringFluid}. In Ref. \cite{Schubring:2014iwa, Schubring:2014ena} it was shown that iMHD is an example of the string fluid whose equations of motion can be obtained by varying the action of Eqs. \eqref{MHD_Lagrangian}. Our work here can thus be viewed as the initiation of an effort to explore the solution space of theories described by action \eqref{SchubringVanchurin_Lagrangian}. 

{\bf Stress Tensors} \qquad By varying the actions \eqref{MHD_Lagrangian_Plasma} and \eqref{MHD_Lagrangian_EM} with respect to the metric, i.e., $g_{\alpha\beta} \to g_{\alpha\beta} + \delta g_{\alpha\beta}$, the stress-energy tensors $T^{\alpha\beta}$ of the plasma and that of the electromagnetic fields can be read off as the coefficients of $-(\sg/2) \sqrt{|g|} \delta g_{\alpha\beta}$. They are
\begin{align}
\label{StressTensor_Plasma}
T[\text{Plasma}]^{\alpha\beta} 
&= \sg g^{\alpha\beta} \left( \rho_0\left[ \sg n^2/2\right] - \rho_0'\left[ \sg n^2/2\right] \sg n^2 \right) 
	+ \rho_0'\left[ \sg  n^2/2\right] n^\alpha n^\beta ,  \\
\label{StressTensor_EM}
T[\text{EM}]^{\alpha\beta} 	
&= \sg \left( - F^{\alpha\sigma} F^{\beta}_{\phantom{\beta}\sigma} + \frac{1}{4} g^{\alpha\beta} F^{\sigma\kappa} F_{\sigma\kappa} \right) .
\end{align}
The total iMHD stress tensor is, of course, their sum
\begin{align}
T[\text{Total}]^{\mu\nu} = T[\text{Plasma}]^{\mu\nu} + T[\text{EM}]^{\mu\nu} .
\end{align}
When the plasma current $n^\mu$ is timelike, i.e., $\sg n^2 > 0$, its stress tensor in eq. \eqref{StressTensor_Plasma} takes on the perfect fluid form
\begin{align}
T[\text{Plasma}]^{\alpha\beta}  = (\rho + p) U^\alpha U^\beta - p \ \sg g^{\alpha\beta}, \qquad\qquad U^\alpha \equiv n^\alpha/\sqrt{|n^2|} ,
\end{align}
where the energy and pressure densities are
\begin{align}
\rho 	&= \rho_0\left[|n^2|/2\right], \\
p		&= -\rho_0\left[|n^2|/2\right] + \rho_0'\left[|n^2|/2\right] |n^2| .
\end{align}
This is the form of the stress tensor derived in \cite{Schubring:2014ena}. However, for the solutions obtained in this paper $n^\mu$ will be null. In such a case, and assuming 
\begin{align}
\lim_{n^2 \to 0} \rho_0'\left[ \sg n^2/2\right] \sg n^2 = 0,
\end{align}
the stress tensor becomes
\begin{align}
\label{StressTensor_Plasma_NullCase}
T[\text{Plasma}]^{\alpha\beta} = \sg \rho_0\left[0\right] g^{\alpha\beta} + \rho_0'\left[0\right] n^\alpha n^\beta .
\end{align}
Furthermore, if we choose a Lagrangian density for the plasma such that its first derivative vanishes at the origin, i.e., $\rho_0 '[0]=0$, then eq. \eqref{StressTensor_Plasma_NullCase} informs us the plasma stress-energy tensor would then take a cosmological-constant-like form:
\begin{align}
\label{StressTensor_Plasma_CCForm}
T[\text{Plasma}]^{\alpha}_{\phantom{\alpha}\beta} = (\sg \rho_0\left[0\right]) \delta^{\alpha}_{\phantom{\alpha}\beta} .
\end{align}
{\bf iMHD equations} \qquad The iMHD action principle in eq. \eqref{MHD_Lagrangian} leads to Maxwell's equations
\begin{align}
\label{EOM1}
\nabla_\sigma \Phi^1 \nabla_\mu F^{\mu\sigma} &= \nabla_\beta \Phi^1 \nabla_\gamma \Phi^3 P^{\beta\gamma}, \\
\label{EOM2}
\nabla_\sigma \Phi^2 \nabla_\mu F^{\mu\sigma} &= \nabla_\beta \Phi^2 \nabla_\gamma \Phi^3 P^{\beta\gamma},
\end{align}
and the plasma equation
\begin{align}
\label{EOM3}
0 &= F_{\beta\gamma} P^{\beta\gamma} ,
\end{align}
with 
\begin{align}
\label{Rank2P}
P^{\beta\gamma} 
&\equiv - \nabla_\alpha \left( \sg \cdot \rho_0'\left[\sg n^2/2\right]\cdot n_\mu \widetilde{\epsilon}^{\mu \alpha\beta\gamma} \right)  \nonumber\\
&= \nabla_\alpha \left( \sg \cdot \rho_0'\left[\sg n^2/2\right] \cdot \nabla^{[\alpha} \Phi_1 \nabla^\beta \Phi_2 \nabla^{\gamma]} \Phi_3 \right) .
\end{align}
The prime indicates a derivative with respect to the argument. We remark in passing that, because the Schubring-Vanchurin class of theories in eq. \eqref{SchubringVanchurin_Lagrangian} are invariant under additive shifts, $\Phi^\text{I} \to \Phi^\text{I} + $ constant, equations \eqref{EOM1}--\eqref{EOM3} can be expressed as the conservation of three Noether currents constructed from the Lagrangian densities in eq. \eqref{MHD_Lagrangian},
\begin{align}
\nabla_\sigma \mathcal{J}^\sigma_{\text{I}} = 0, \qquad\qquad 
\mathcal{J}^\sigma_{\text{I}} \equiv \frac{\partial \left( - \rho_0\left[ \sg n^2/2 \right] - \varphi^2/2 \right)}{\partial\left( \nabla_\sigma \Phi^\text{I} \right)}, 
\qquad\qquad \text{I} \in \{ 1,2,3 \} .
\end{align}
{\bf Force-Free Limit} \qquad By using Maxwell's equations
\begin{align}
\label{EMCurrent}
\nabla_\mu F^{\mu\nu} 				&= \frac{\partial_\mu \left( \sqrt{|g|} g^{\mu\alpha} g^{\nu\beta} F_{\alpha\beta} \right)}{\sqrt{|g|}} = J^\nu, \\
\nabla_{[\alpha} F_{\beta\gamma]} 	&= 0 ,
\end{align}
where $J$ is the electromagnetic current\footnote{Within iMHD, where there is no externally prescribed electromagnetic current $J^\nu$, eq. \eqref{EMCurrent} is to be viewed as the definition of $J^\nu$.}, the divergence of the electromagnetic stress tensor in eq. \eqref{StressTensor_EM} is
\begin{align}
\label{StressTensor_EM_div}
\nabla_\mu T[\text{EM}]^{\mu\nu} = \sg \ J_\mu F^{\mu\nu} . 
\end{align}
The force-free condition holds whenever the electromagnetic current becomes orthogonal to the Maxwell field strength,
\begin{align}
\label{ForceFreeCondition}
J_\mu F^{\mu\nu} = 0.
\end{align}
Referring to eq. \eqref{StressTensor_EM_div}, we see that the force-free condition of eq. \eqref{ForceFreeCondition} is thus equivalent to the conservation of the electromagnetic stress-energy tensor $\nabla_\mu T$[EM]$^{\mu\nu} = 0$.  When electromagnetic fields are not the only matter present in the system, such as the case for iMHD, this is not a trivial requirement, since it is usually the total stress tensor that is conserved $\nabla_\mu T$[Total]$^{\mu\nu} = 0$. On the other hand, even though $\rho_0$ in eq. \eqref{MHD_Lagrangian} is usually set to zero when ``Force-Free Electrodynamics" (FFE) is discussed in the literature -- as we shall proceed to show explicitly, when the force-free condition in eq. \eqref{ForceFreeCondition} is obeyed, this does not necessarily imply there is no other matter present.

In \S \eqref{Section_Solutions}, we will delineate two families of analytic iMHD solutions that, despite the presence of a non-trivial plasma, continues to obey the force-free condition. To this end, we seek a null plasma current so that the plasma stress-energy tensor (and thus that of the electromagnetic fields) becomes separately conserved because $T$[Plasma]$^{\mu\nu}$ takes a cosmological-constant-like form in eq. \eqref{StressTensor_Plasma_CCForm}, due to the choice in eq. \eqref{rho0prime}. In \S \eqref{Section_SelfConsistent} we highlight a special case that allows us to obtain self-consistent solutions to the Einstein-iMHD equations. We then close in \S \eqref{Section_Outlook} with comments on possible future research directions.

\section{Two Analytic iMHD Solutions}
\label{Section_Solutions}
{\bf Spacetime geometry} \qquad We will work with the class of 4D curved spacetimes containing a free function $f$,
\begin{align}
\label{g}
\dd s^2 = \sg \left\{ f[u^\pm,r,\theta,\phi] (\dd u^\pm)^2 \pm 2 \dd u^\pm \dd r - g_{\mathfrak{A}\mathfrak{B}} \dd \psi^\mathfrak{A} \dd \psi^\mathfrak{B} \right\} ,
\end{align}
where\footnote{The $\pm$ in eq. \eqref{g} refers to either the ingoing $u^-$ or outgoing $u^+$ null coordinate; for the rest of the paper, every time there is a choice of signs, the upper one would apply if $u^+$ is being used and the lower one if $u^-$ is employed instead.} the metric on the $2-$sphere of radius $r$ is
\begin{align}
\label{g_2Sphere}
g_{\mathfrak{A}\mathfrak{B}} \dd \psi^{\mathfrak{A}} \dd \psi^{\mathfrak{B}} = r^2 \left( \dd\theta^2 + \sin^2 \theta \dd\phi^2 \right), \qquad\qquad 
\psi^\mathfrak{A} \equiv (\theta,\phi), \qquad\qquad
\mathfrak{A}, \mathfrak{B} \in \left\{ 2,3 \right\} .
\end{align}
Because $f$ is arbitrary, the metric in eq. \eqref{g} includes the Vaidya-(anti-)de Sitter class of spacetimes describing a dark energy dominated cosmology with null matter accreting onto a central black hole, where
\begin{align}
\label{Vaidya}
f_\text{V(A)dS}[u^\pm,r] = 1 - \frac{\Lambda}{3} r^2 - \frac{2 \GN M[u^\pm]}{r} .
\end{align}
This can be reduced to the Schwarzschild black hole, by setting the cosmological constant $\Lambda$ to zero and the mass $M$ to a constant.

{\bf Vanishing of plasma energy density's first derivative} \qquad Notice from eq. \eqref{Rank2P} that the right hand side of equations \eqref{EOM1}--\eqref{EOM3} will vanish if we can arrange for the plasma current to be null ($n^2=0$) and for the first derivative of the plasma energy density to vanish there, namely
\begin{align}
\label{rho0prime}
\rho_0'[0] = 0 .
\end{align}
If we can also find $\Phi^{1,2}$ such that they obey the force-free condition of eq. \eqref{ForceFreeCondition}, we would then have solved the full set of iMHD equations \eqref{EOM1}--\eqref{EOM3}. This scenario will play out for the following two families of iMHD solutions.

{\bf Brannan-Gralla-Jacobson-iMHD solution} \qquad The first set of solutions we have found is a direct generalization of that found by Brennan, Gralla and Jacobson \cite{Brennan:2013kea} (see also \cite{Gralla:2014yja}) for FFE, with $\rho_0=0$ in eq. \eqref{MHD_Lagrangian}, to those of the full iMHD system in equations \eqref{EOM1} through \eqref{EOM3}. Their solution, involving only our $\Phi^{1,2}$, consists of
\begin{align}
\label{BGJ_IofII}
\Phi^1 = \zeta[u^\pm,\theta,\phi], \qquad\qquad \Phi^2 = u^\pm .
\end{align}
In addition, we will assume eq. \eqref{rho0prime} holds, and proceed to make $n^2=0$ -- thereby satisfying all three equations \eqref{EOM1} through \eqref{EOM3} -- by putting
\begin{align}
\label{BGJ_IIofII}
\Phi^3 = Z[u^\pm,\theta,\phi] .
\end{align}
{\bf Axis-symmetric Maxwell tensor} \qquad The second set of solutions involve assuming cylindrical symmetry of $\Phi^1$ but now allow $\Phi^2$ to depend on the altitude angle $\theta$,
\begin{align}
\label{AxiallySymmetric}
\Phi^1 = X[u^\pm,\theta], \qquad\qquad \Phi^2 = Y[u^\pm,\theta], \qquad\qquad \Phi^3 = Z[u^\pm,\theta,\phi] .
\end{align}
We continue to assume eq. \eqref{rho0prime}. 

For both solutions in equations \eqref{BGJ_IofII}--\eqref{BGJ_IIofII} and \eqref{AxiallySymmetric},
\begin{align}
F_{\alpha\beta} F^{\alpha\beta} = 0 = \widetilde{\epsilon}^{\mu\nu \alpha\beta} F_{\mu\nu} F_{\alpha\beta} .
\end{align}
In the following subsections, we shall describe the derivation of these solutions. Following that, in \S \eqref{Section_SelfConsistent}, we will find a special case of eq. \eqref{AxiallySymmetric} such that the full Einstein-iMHD system may be solved simultaneously.

\subsection{Brennan-Gralla-Jacobson solution generalized to iMHD}
\label{Section_BGJ}
We start with the solutions in eq. \eqref{BGJ_IofII} and \eqref{BGJ_IIofII}; when $\rho_0 \to 0$ in eq. \eqref{MHD_Lagrangian} and $\Phi^3$ is neglected, the $\Phi^{1,2}$ are the solutions in \cite{Brennan:2013kea}. Suppose for the moment that the $Z$ in eq. \eqref{BGJ_IIofII} depended on $r$ as well,
\begin{align}
\Phi^3 = Z[u^\pm,r,\theta,\phi] .
\end{align}
One would find that the plasma current is 
\begin{align}
\label{BGJ_PlasmaCurrent}
n^\mu = \frac{\csc[\theta]}{r^2} \left(
0, \partial_{[\phi} \zeta \partial_{\theta]} Z, -\partial_{\phi} \zeta \partial_r Z, \partial_\theta \zeta \partial_r Z
\right) ,
\end{align}
which indicates it is either null or spacelike
\begin{align}
n^2 = -\sg (\partial_r Z)^2 g^{\mathfrak{A}\mathfrak{B}} \partial_\mathfrak{A} \zeta \partial_\mathfrak{B} \zeta .
\end{align}
To make $n^2=0$, one can choose $Z$ to be $r$-independent; or,
\begin{align}
0 
= g^{\mathfrak{A}\mathfrak{B}} \partial_{\mathfrak{A}} \zeta \partial_{\mathfrak{B}} \zeta 
= \frac{(\partial_\theta \zeta)^2 + \csc^2[\theta] (\partial_\phi\zeta)^2}{r^2} .
\end{align}
But since this latter choice would be setting to zero the sum of the two squares $(\partial_\theta \zeta)^2$ and $\csc^2[\theta] (\partial_\phi\zeta)^2$, for it to hold for all angles, $\zeta$ must therefore be independent of both angular coordinates -- this would mean the plasma current in eq. \eqref{BGJ_PlasmaCurrent} becomes trivial. Hence, to have a non-zero null plasma current while keeping the Brennan-Gralla-Jacobson solutions in eq. \eqref{BGJ_IofII}, we must let $Z$ be independent of $r$. A direct calculation will show that at this point, out of the 3 iMHD equations, only eq. \eqref{EOM1} is non-trivial:
\begin{align}
0 = \pm \frac{2 \csc ^2[\theta] \rho_0'[0]}{r^5} \left( \partial_{[\phi} \zeta \partial_{\theta]} Z \right)^2 .
\end{align}
We see that this equation is satisfied if $\rho_0'[0]$ is chosen to be zero. ($Z$ cannot depend on spacetime through $u^\pm$ alone, because this would set to zero the plasma current in eq. \eqref{BGJ_PlasmaCurrent}.)

The electromagnetic current $J^\nu$ is non-zero only in its $r$-component, which in turn can be expressed as ($\mp$ times) the covariant Laplacian, with respect to the $2-$sphere metric, acting on $\zeta$:
\begin{align}
J^\nu \equiv \nabla_\mu F^{\mu\nu} = \mp \left( 0,g^{\mathfrak{A}\mathfrak{B}} \nabla_{\mathfrak{A}} \nabla_{\mathfrak{B}} \zeta ,0,0 \right) .
\end{align}
Since $g_{rr} = 0$, observe that $J^\nu$ is null. The stress-energy tensor of the plasma takes the form in eq. \eqref{StressTensor_Plasma_CCForm}; whereas, the only non-zero component of the Maxwell  stress-energy tensor \eqref{StressTensor_EM} is
\begin{align}
T[\text{EM}]_{\pm \pm} = g^{\mathfrak{A}\mathfrak{B}} \partial_{\mathfrak{A}} \zeta \partial_{\mathfrak{B}} \zeta .
\end{align}
Consistency checks can be made by ensuring that the FFE condition in eq. \eqref{ForceFreeCondition} is satisfied; and the divergence of the plasma and electromagnetic stress-energy tensors are separately zero ($\nabla_\mu T[\text{Plasma}]^{\mu\nu} = \nabla_\mu T[\text{EM}]^{\mu\nu} = 0$). 

\subsection{iMHD Solution With Axially Symmetric Maxwell Tensor}
\label{Section_AxiallySymmetric}
Now, we will allow $\Phi^2 = Y[u^\pm,\theta]$ to pick up a $\theta$ dependence at the cost of assuming cylindrical symmetry for $\Phi^1 = X[u^\pm,\theta]$. This ansatz for $\Phi^{1,2}$ maintains the force-free condition in eq. \eqref{ForceFreeCondition}. Following the previous subsection, if we suppose for the moment that the $Z$ in eq. \eqref{AxiallySymmetric} depended on $r$ as well,
\begin{align}
\Phi^3 = Z[u^\pm,r,\theta,\phi] .
\end{align}
The corresponding plasma current is
\begin{align}
\label{AxiallySymmetric_PlasmaCurrent}
n^\mu = \frac{\csc[\theta]}{r^2} \partial_{[\pm} X \partial_{\theta]} Y 
\left( 0, \partial_\phi Z, 0, -\partial_r Z \right)  ,
\end{align}
such that one would find that it is either null or spacelike
\begin{align}
n^2 = -\sg \frac{(\partial_r Z)^2}{r^2} \left( \partial_{[\pm} X \partial_{\theta]} Y \right)^2 .
\end{align}
To make $n^2=0$, one can choose $Z$ to be $r$-independent; or,
\begin{align}
0 = \partial_{[\pm} X \partial_{\theta]} Y .
\end{align}
However, this latter choice would render the entire plasma current in eq. \eqref{AxiallySymmetric_PlasmaCurrent} trivial. Hence, to have a non-zero null plasma current while maintaining the form of the $\Phi^{1,2}$ solutions in eq. \eqref{AxiallySymmetric}, we must let $Z$ be independent of $r$. At this point, a direct calculation would show, out of the 3 iMHD equations, only eq. \eqref{EOM1} and \eqref{EOM2} are non-trivial:
\begin{align}
0 &= \pm \frac{2 \csc ^2[\theta] \rho_0'[0]}{r^5} \partial_\theta X \left( \partial_{[\theta} X \partial_{\pm]} Y \right) (\partial_\phi Z)^2  , \\
0 &= \pm \frac{2 \csc ^2[\theta] \rho_0'[0]}{r^5} \partial_\theta Y \left( \partial_{[\theta} X \partial_{\pm]} Y \right) (\partial_\phi Z)^2 .
\end{align}
We see that this pair of equations are satisfied if $\rho_0'[0]$ is chosen to be zero. ($Z$ cannot be $\phi$-independent because that would render the plasma current in eq. \eqref{AxiallySymmetric_PlasmaCurrent} trivial.)

The only non-zero component of the electromagnetic current is
\begin{align}
J^r = \nabla_\mu F^{\mu r} 
&=  \pm r^{-2} \Big(
\partial_u X \partial_\theta^2 Y - \partial_\theta X \partial_u \partial_\theta Y \\
&\qquad\qquad
- \partial_u Y \left ( \partial_\theta^2 X + \cot[\theta] \partial_\theta X \right )
+ \partial_\theta Y \left ( \partial_u \partial_\theta X + \cot[\theta] \partial_u X \right)
\Big) . \nonumber
\end{align}
Because $g_{rr} = 0$, $J^\nu$ is null. The stress-energy tensor of the plasma takes the form in eq. \eqref{StressTensor_Plasma_CCForm}; whereas, the only non-zero component of the Maxwell stress-energy tensor \eqref{StressTensor_EM}  is
\begin{align}
T[\text{EM}]_{\pm \pm} = \left( \frac{\partial_{[\pm} X \partial_{\theta]} Y}{r} \right)^2 .
\end{align}
For consistency checks, we have ensured the divergence of the plasma and electromagnetic stress-energy tensors are separately zero ($\nabla_\mu T[\text{Plasma}]^{\mu\nu} = \nabla_\mu T[\text{EM}]^{\mu\nu} = 0$). 

\section{A Self-Consistent Einstein-iMHD Solution}
\label{Section_SelfConsistent}
In this section we wish to present a self-consistent solution to the Einstein-iMHD equations. We start with the axially-symmetric-Maxwell ansatz of eq. \eqref{AxiallySymmetric} and the metric in eq. \eqref{g}. Since the iMHD equations are already satisfied, we shall focus on Einstein's:\footnote{To set conventions, we record here that our Ricci tensor is $R_{\beta\nu} \equiv \partial_{[\mu} \Gamma^\mu_{\phantom{\alpha}\nu]\beta} + \Gamma^\mu_{\phantom{\mu}\sigma [\mu} \Gamma^\sigma_{\phantom{\sigma}\nu]\beta}$, while the Christoffel symbols themselves are $\Gamma^\mu_{\phantom{\mu}\alpha \beta} \equiv (1/2) g^{\mu\sigma} ( \partial_\alpha g_{\beta\sigma} + \partial_\beta g_{\alpha\sigma} - \partial_\sigma g_{\alpha\beta} )$. The Einstein tensor is $G_{\beta\nu} \equiv R_{\beta\nu} - (g_{\beta\nu}/2) \mathcal{R}$, where the Ricci scalar is $\mathcal{R} \equiv g^{\beta\nu} R_{\beta\nu}$.}
\begin{align}
\label{Einstein}
G[g]^\mu_{\phantom{\mu}\nu} - \Lambda_\text{cc} \cdot \sg \delta^\mu_{\phantom{\mu}\nu} = 8\pi \GN T[\text{Total}]^\mu_{\phantom{\mu}\nu} .
\end{align}
The $00$ component hands us a first-order-in-$r$ equation for $f$,
\begin{align}
\frac{1-\partial_r(rf)}{r^2} = \Lambda_\text{cc} + 8 \pi \GN \rho_0[0],
\end{align}
which may be readily integrated to yield
\begin{align}
f[u^\pm,r,\theta,\phi] = 1 - \frac{\Lambda_\text{cc} + 8\pi \GN \rho_0[0]}{3} r^2 + \frac{\chi[u^\pm,\theta,\phi]}{r} .
\end{align}
The $r\theta$ and $r\phi$ components of eq. \eqref{Einstein} then tell us $\chi$ needs, in fact, to be independent of the angular coordinates.
\begin{align}
G[g]^r_{\phantom{r}\theta} 	&= - \sg \frac{\partial_\theta \chi}{2 r^2} = 8 \pi \GN T[\text{Total}]^r_{\phantom{r}\theta} = 0 , \\
G[g]^r_{\phantom{r}\phi} 	&= - \sg \frac{\partial_\phi \chi}{2 r^2} 	= 8 \pi \GN T[\text{Total}]^r_{\phantom{r}\phi} = 0 .
\end{align}
At this point, we re-define $\chi \equiv -2\GN M[u^\pm]$, i.e.,
\begin{align}
\label{SelfConsistent_f}
f[u^\pm,r,\theta,\phi] = 1 - \frac{\Lambda_\text{cc} + 8\pi \GN \rho_0[0]}{3} r^2 - \frac{2 \GN M[u^\pm]}{r} .
\end{align}
Comparison with eq. \eqref{Vaidya} shows we may identify the ``effective cosmological constant" as
\begin{align}
\label{SelfConsistent_Lambda_eff}
\Lambda = \Lambda_\text{cc} + 8 \pi \GN \rho_0[0] .
\end{align}
Finally, by examining the $r \pm$ components of eq. \eqref{Einstein}, we find that Einstein's equations have been reduced to
\begin{align}
\label{VaidyaMass_Eqn}
M'[u^\pm] = \mp 4\pi \left( \partial_{[\pm} X \partial_{\theta]} Y \right)^2 .
\end{align}
Eq. \eqref{VaidyaMass_Eqn} can be integrated to solve for $M$ if and only if $\partial_{[\pm} X \partial_{\theta]} Y$ on the right hand side can be made $\theta$-independent. Let us view $\partial_{[\pm} X \partial_{\theta]} Y$ is the non-trivial component $\mathcal{F}_{\pm \theta} = -\mathcal{F}_{\theta\pm}$ of an antisymmetric tensor in the 2-dimensional space parametrized by $(u^\pm,\theta)$,\footnote{The following argument was inspired by Appendix D of \cite{Gralla:2014yja}, which in turn was based on \cite{Uchida}.} i.e.,
\begin{align}
\mathcal{F}_{\sf AB} 
\equiv \partial_{\sf [ A} X \partial_{\sf B ]} Y, \qquad\qquad 
\partial_{\sf A} \equiv (\partial_\pm,\partial_\theta) .
\end{align}
(This $\mathcal{F}_{\sf AB}$ is of course the 4D Maxwell tensor restricted to the 2D $(u^\pm,\theta)$--plane.) Furthermore, let $\xi^{\sf A} \partial_{\sf A} = \partial_\theta$. Then the requirement that $\mathcal{F}_{\pm \theta}$ be $\theta$-independent can be phrased in the following covariant form:
\begin{align}
\label{d(Fxi)IsZero}
\partial_{\sf [A} \left( \mathcal{F}_{\sf B]C} \xi^{\sf C} \right) = 0 .
\end{align}
To see this we note that the only non-trivial components are ${\sf AB} = \pm \theta$ and ${\sf AB} = \theta \pm$; moreover, since $\mathcal{F}_{\theta\theta} = 0$, by setting say ${\sf AB} = \theta \pm$,
\begin{align}
\partial_{\theta} \mathcal{F}_{\pm \theta} = 0 .
\end{align}
In form notation, eq. \eqref{d(Fxi)IsZero} is denoted as
\begin{align}
\dd \left( \mathcal{F} \cdot \xi \right) = 0 ;
\end{align}
this in turn implies $\mathcal{F} \cdot \xi$ is a gradient of some function $H[u^\pm,\theta]$:
\begin{align}
\mathcal{F}_{\sf BC} \xi^{\sf C} = \mathcal{F}_{{\sf B}\theta} = \partial_{\sf B} H .
\end{align}
When ${\sf B}=\theta$, this equation tells us $H$ is actually independent of $\theta$: $0 = \partial_\theta H$. When ${\sf B}=\pm$,
\begin{align}
\label{SelfConsistent_dXdY}
\partial_{\pm} X \partial_\theta Y - \partial_{\pm} Y \partial_\theta X = H'[u^\pm] . 
\end{align}
If we proceed to perform a change-of-variables $\partial_\pm = H' \partial_H$,
\begin{align}
\label{SelfConsistent_XYEqn}
\partial_{[H} \left( X\left[ H,\theta \right] \partial_{\theta]} Y\left[H,\theta\right] \right) = 1.
\end{align}
Again, if we view the left hand side as the non-trivial component of the 2D antisymmetric tensor
\begin{align}
\mathcal{F}_{{\sf A}'{\sf B}'} \equiv \partial_{\sf [ A} X \partial_{\sf B ]} Y, \qquad\qquad \partial_{\sf A} = (\partial_H,\partial_\theta) ,
\end{align}
the $X \dd Y$ can be viewed as the corresponding gauge potential. Eq. \eqref{SelfConsistent_XYEqn} translates to
\begin{align}
\dd \left( X \dd Y \right) = \dd \left( H \dd \theta \right),
\end{align}
which implies the gauge potential is 
\begin{align}
\label{SelfConsistent_GaugePotential}
X \dd Y &= H \dd \theta + \dd K ,
\end{align}
where $K$ is some arbitrary function. Referring to the iMHD equations \eqref{EOM1}--\eqref{EOM3} (and \eqref{Rank2P}), as well as the definition of the plasma current in eq. \eqref{PlasmaCurrent}, let us keep in mind that it is only the components of $F = \dd (X \dd Y)$ that are relevant, not the individual $X$ and $Y$ themselves. Since $\dd^2 K = 0$ anyway, we ``choose a gauge" and set $K=$ constant. Following that, we proceed to expand the left hand side of eq. \eqref{SelfConsistent_GaugePotential} in the basis forms $\dd H$ and $\dd \theta$, 
\begin{align}
\label{SelfConsistent_GaugePotential_BasisForms}
X (\partial_H Y \dd H + \partial_\theta Y \dd \theta) 	&= H \dd \theta .
\end{align}
Equating the coefficient of $\dd H$ on both sides of eq. \eqref{SelfConsistent_GaugePotential_BasisForms} then informs us that $Y$ itself must be $H$-independent:
\begin{align}
\label{SelfConsistent_YSoln}
\Phi^2 =Y[\theta] .
\end{align}
Equating the coefficient of $\dd \theta$ on both sides of eq. \eqref{SelfConsistent_GaugePotential_BasisForms}, and taking eq. \eqref{SelfConsistent_YSoln} into account, directs us to
\begin{align}
\label{SelfConsistent_XSoln}
\Phi^1 = X[u^\pm] = \frac{H[u^\pm]}{Y'[\theta]} .
\end{align}
Our arguments have determined the form of $X$ and $Y$ that would give us the most general $\theta$-independent expression for $\partial_{[\pm} X \partial_{\theta]} Y$. From equations \eqref{SelfConsistent_YSoln} and \eqref{SelfConsistent_XSoln}, we may check explicitly that eq. \eqref{SelfConsistent_dXdY} is recovered; whereas, the third scalar field remains as
\begin{align}
\label{SelfConsistent_ZSoln}
\Phi^3 = Z[u^\pm,\theta,\phi] .
\end{align}
We have therefore determined that
\begin{align}
\label{VaidyaMass_Eqn_HPrime}
M'[u^\pm] = \mp 4\pi H'[u^\pm]^2 .
\end{align}
To sum: the scalar fields in eq. \eqref{SelfConsistent_YSoln}--\eqref{SelfConsistent_ZSoln} not only satisfy the iMHD equations \eqref{EOM1}--\eqref{EOM3} (as well as the force-free condition eq. \eqref{ForceFreeCondition}); they also satisfy Einstein's equations in eq. \eqref{Einstein}, sourcing the metric in eq. \eqref{g} with the particular Vaidya-(anti-)de Sitter $f$ in eq. \eqref{SelfConsistent_f} -- provided eq. \eqref{VaidyaMass_Eqn_HPrime} holds for the ``mass function" $M$. Notice the effective cosmological constant in eq. \eqref{SelfConsistent_Lambda_eff} receives contributions only from the plasma; whereas the mass $M$, through eq. \eqref{VaidyaMass_Eqn_HPrime}, does so only from electromagnetism.

To be sure our solution here is not trivial, we record here the various physical tensors of the setup. The only non-zero component of the plasma current is
\begin{align}
n^r = \frac{\csc[\theta] H'[u^\pm] \partial_\phi Z}{r^2} .
\end{align}
The only non-zero component of the electromagnetic current is
\begin{align}
J^r = \pm \frac{\cot[\theta] H'[u^\pm]}{r^2} .
\end{align}
The only non-zero component of the Maxwell tensor is
\begin{align}
F_{\pm \theta} = -F_{\theta \pm} = H'[u^\pm] .
\end{align}
Next, the plasma stress-tensor is
\begin{align}
\label{SelfConsistent_Plasma_StressTensor}
T[\text{Plasma}]_{\mu\nu} = (\sg \rho_0[0]) \cdot g_{\mu\nu} ;
\end{align}
while the only non-zero component of the electromagnetic one is
\begin{align}
\label{SelfConsistent_EM_StressTensor}
T[\text{EM}]_{\pm \pm} = \left( \frac{H'[u^\pm]}{r} \right)^2 .
\end{align}

\section{Summary and Outlook}
\label{Section_Outlook}
We have found two families of iMHD solutions, one in eq. \eqref{BGJ_IofII}--\eqref{BGJ_IIofII} and another in eq. \eqref{AxiallySymmetric}, in the curved background geometry of eq. \eqref{g}. Additionally, we discovered that the scalar fields in equations \eqref{SelfConsistent_YSoln}, \eqref{SelfConsistent_XSoln}, and \eqref{SelfConsistent_ZSoln} together with the Vaidya-(anti-)de Sitter metric for which $f$ in eq. \eqref{SelfConsistent_f} is employed in eq. \eqref{g}, simultaneously solve the equations of iMHD \eqref{EOM1}--\eqref{EOM3} and of Einstein's \eqref{Einstein}, provided the mass function is subject to eq. \eqref{VaidyaMass_Eqn}. Because all these solutions assume eq. \eqref{rho0prime}, they still maintain the force-free condition of eq. \eqref{ForceFreeCondition} despite the presence of a plasma.

We close with thoughts on possible next steps to take. For astrophysical applications, it would be important to seek iMHD solutions in axially symmetric geometries or Kerr black hole backgrounds, by perhaps once again extending the known FFE solutions. We also have not studied the stability of the iMHD solutions in this paper. Finally, we believe it is of physical interest to move away from the ideal MHD limit, and include effects from dissipation \cite{Schubring:2014iwa}.

\section{Acknowledgments}
\label{Section_Acknowledgments}

We thank Daniel Schubring for discussions. Much of the analytic work in this paper was done with {\sf Mathematica} \cite{Mathematica}.

\end{document}